\documentclass[conference]{IEEEtran}

\usepackage{multirow}
\usepackage{tikz}
\usetikzlibrary{arrows.meta,positioning,calc,fit}
\usepackage{cite}
\usepackage{amsmath,amssymb,amsfonts}
\usepackage{algorithmic}
\usepackage{graphicx}
\usepackage{textcomp}
\usepackage{import}
\usepackage{transparent}
\usepackage{balance}
\usepackage{stfloats}
\usepackage{balance}

\usepackage{xcolor}
\usepackage[utf8]{inputenc}
\usepackage{textgreek}
\usepackage{tabularx}
\usepackage[normalem]{ulem}
\usepackage{booktabs}
\usepackage{arydshln}
\usepackage[hidelinks]{hyperref}
\IEEEoverridecommandlockouts

\begin{document}

\title{Conditional What-If Generation for Radio Access Network Energy under Rare and Unseen Contexts}
\newcommand{\papertitle}{Conditional What-If Generation for Radio Access Network Energy under Rare and Unseen Contexts}

\newcommand{\runningtitle}{Conditional RAN Energy Generation}

\hypersetup{
    pdftitle={\papertitle},
    pdfauthor={Kevin Wardakhan, Marie Line Alberi-Morel, Jeremie Leguay, Afef Feki, Adrien Sardi, Sebastien Archeveque, Claudiu Mihailescu},
    pdfsubject={\runningtitle}
}

\author{
\IEEEauthorblockN{
Kevin Wardakhan$^{1}$,
Marie Line Alberi-Morel$^{1,*}$\thanks{\textsuperscript{*}Corresponding author: \href{mailto:marie_line.alberi-morel@nokia-bell-labs.com}{marie\_line.alberi-morel@nokia-bell-labs.com}.},
J\'er\'emie Leguay$^{1}$,
Afef Feki$^{2}$,
Adrien Sardi$^{1,3}$, \\
S\'ebastien Archeveque$^{2}$,
Claudiu Mihailescu$^{2}$
}
\IEEEauthorblockA{
$^{1}$Nokia Bell Labs, $^{2}$Nokia Mobile Infrastructure, $^{3}$Inria, Université Côte d'Azur, France
}
}

\maketitle

\pagestyle{plain}
\thispagestyle{plain}

\begin{abstract}
Energy-saving mechanisms in Radio Access Networks (RANs) must balance energy reduction and network performance under varying traffic conditions. Conditional generative models provide a framework for what-if analysis by generating  plausible power consumption trajectories for different radio configurations and operating scenarios. This capability helps operators evaluate alternative energy-saving policies before deployment. However, collecting representative measurements is costly and operationally constrained, so historical data cover a limited set of configurations and operating conditions, leaving some contexts poorly represented or entirely unseen during training. In this paper, we study conditional generation of RAN power consumption trajectories under decreasing training coverage, from well-represented to rare and unseen configurations. We first introduce a context-normalized conditional generator combining temporally aligned conditioning with configuration-dependent target normalization. This approach improves conditional generation for well-represented configurations, but becomes less reliable as training exposure decreases because target power consumption becomes harder to estimate. We therefore introduce a prediction-based generative framework that decomposes power consumption into global prediction, residual level and scale correction, and temporal residual generation. Experiments on measurements from a commercial 5G RAN and a public dataset show that hardware metadata provides transferable information about residual scale, while limited target observations at inference  improve conditional generation through residual-level correction, without retraining. An ablation study confirms that residual-level correction provides most of the gain. Both approaches improve over state-of-the-art methods for well-represented configurations, while the prediction-based approach also generalizes to rare and unseen contexts and, with only one target observation, reduces the median MAE by about $50\%$ compared to TimeWeaver.
\end{abstract}


\section{Introduction}

Energy optimization has become a first-order concern in mobile networks, as sustained traffic growth and network densification increasingly offset the efficiency gains brought by successive generations such as 5G/6G~\cite{lopez2022survey}. At the same time, operators face rising energy costs and increasingly stringent sustainability objectives. The Radio Access Network (RAN), and base stations in particular, accounts for a major share of energy consumption, with power amplifiers (PAs) responsible for about 64\% of the radio-site power consumption~\cite{maggi2023energy}. Consequently, mechanisms such as carrier shutdown, MIMO muting and antenna deactivation have become important levers for reducing both operational expenditure and carbon footprint. Despite advances in massive MIMO, beamforming, and lean carrier design, substantial system-level savings still depend on dynamically adapting active radio resources to varying traffic demand. 

However, configuring these mechanisms is challenging because the trade-off between energy savings and network performance depends strongly on the operating conditions. Aggressive shutdown decisions concentrate traffic on fewer active resources and can degrade KPIs such as throughput, accessibility, or mobility, whereas conservative policies leave much of the potential energy saving untapped. Existing approaches therefore rely on carefully tuned control parameters, such as load thresholds that determine when and to what extent energy-saving mechanisms are activated or deactivated. For instance,~\cite{maggi2023energy} optimizes carrier-shutdown thresholds through closed-loop Bayesian learning and demonstrates an 11\% reduction in base station power consumption while satisfying predefined KPI constraints in a live network. Yet evaluating candidate configurations directly in a live RAN is inherently costly and potentially risky, particularly under traffic conditions for which little operational experience exists. This motivates the development of what-if analysis tools able to estimate how the network would behave under alternative radio configurations and hypothetical operating conditions before they are effectively deployed, and more broadly calls for AI-driven tools that can support safe and adaptive RAN optimization~\cite{sthankiya2024survey,li2025generative}.

Data-driven power consumption models provide a natural starting point for such tools by learning the relationship between hardware configuration, traffic conditions, energy-saving mechanisms, and power consumption. For example,~\cite{piovesan2022machine} trains a neural model across base stations and demonstrates that knowledge can generalize across equipment configurations. Such predictive models are well suited to estimating power at a given operating point, but what-if analysis can benefit from a richer representation of uncertainty: rather than producing only an expected behavior or pointwise conditional distribution, time-series generative models can sample multiple temporally coherent trajectories, thereby representing the range of plausible network responses to a same requested traffic scenario. Recent conditional generators provide a way to model such trajectory distributions. In particular, TimeWeaver~\cite{narasimhan2024timeweaver} conditions diffusion-based generation on heterogeneous static and time-varying metadata, while CENTS (Context Encoding and Normalizing Time Series Generation)~\cite{fuest2025cents} addresses the complementary challenge of generating power consumption data for rare and previously unseen combinations of contextual variables. However, generating consumption trajectories for sparsely represented radio configurations
and operating conditions requires not only capturing temporal dynamics, but
also estimating the appropriate energy consumption level and variability with limited
historical data.

In this paper, we investigate conditional generation of RAN power consumption trajectories as the amount of training data available for the requested context decreases from well-seen, or well-represented, to rarely seen and unseen contexts. 
We first introduce and evaluate a context-normalized conditional generator that combines TimeWeaver's temporally aligned conditioning with context encoding and configuration-dependent normalization inspired by CENTS. This approach performs well when the requested configuration is well represented during training, but becomes less reliable as training exposure decreases, mainly because estimating the consumption-level  becomes more difficult under rare and unseen configurations.
To mitigate this limitation, we introduce a novel prediction-based generative framework that separates consumption level and variability from temporal residual generation. Instead of estimating power in a single step, it decomposes it into global power prediction, residual level and scale correction, and temporal residual generation. The two approaches are closely related, as the temporal component of the prediction-based framework can reuse the conditional generator introduced above, but applied to standardized residual trajectories rather than directly to power.

The key contributions of this work are two novel approaches for generating plausible power-consumption trajectories for what-if analysis: (i) a direct conditional generation method and (ii) a prediction-based generative framework with residual-correction that requires no retraining for adaptation. We evaluate both approaches on measurements from a commercial 5G RAN and a public dataset~\cite{huawei2023energy}, and conduct an ablation study to assess the contribution of the different components. The results confirm the dominant role of  residual-level correction and show that the prediction-based approach generalizes better to rarely seen and unseen cases. Finally, the latter approach also exhibits superior generalization to unseen operating regimes or configurations that were not encountered during the training.

The remainder of the paper is organized as follows. Sec.~\ref{sec:related} reviews related work. Sec.~\ref{sec:problem} formulates the RAN energy what-if generation problem, describes the datasets, and discusses the context-sparsity challenge. Sec.~\ref{sec:approach} presents the proposed generation and prediction-based correction methods, Sec.~\ref{sec:results} evaluates their performance across different contexts and operating scenarios, and Sec.~\ref{sec:conclusion} concludes the paper.

\section{Related Work}
\label{sec:related}

In this section, we review related work in RAN energy modeling and conditional time-series generation.

\subsection{RAN Energy Modeling and AI-Driven Network Twins}

Generative AI is increasingly considered as an enabling technology for Network Digital Twins (NDTs), where learned models can complement conventional simulation by reproducing network behavior under hypothetical conditions. Recent works identify network behavior generation and what-if analysis as key applications for generative-AI-enabled NDTs~\cite{li2025generative}. Earlier data-driven approaches already explored this principle without generative models. For instance, traffic forecasting and learned performance models have been combined to build digital-twin scenario simulators that evaluate the impact of hypothetical traffic changes on network QoS~\cite{dulas2024data}. These works motivate learned what-if models, but do not address probabilistic generation of RAN energy trajectories under rare or unseen operating contexts.

For RAN energy, existing data-driven approaches are predominantly predictive. Neural power consumption models have been learned from hardware characteristics, traffic conditions, and energy-saving states, showing that data-driven models can provide more accurate representations of complex 5G power consumption behavior than conventional analytical models~\cite{piovesan2022machine}. Multi-carrier extensions further show that a common neural model can capture different base station types and transfer knowledge across equipment configurations~\cite{piovesan2023multicarrier}. More recently, Energy Fingerprint incorporates base station identity through learned embeddings and masked training to capture consumption characteristics of individual radio units while improving generalization~\cite{chen2025energyfingerprint}. These approaches accurately estimate consumption from a given operating state, but remain pointwise predictors. Our work addresses this gap by considering conditional generation as a probabilistic surrogate for RAN energy what-if analysis.

\subsection{Conditional Time-Series Generation}

Conditional time-series generation focuses on generating plausible trajectories conditioned on contextual information. Diffusion models have emerged as a powerful approach for this task. Conditional Score-based Diffusion models for Imputation (CSDI) introduced a diffusion-based framework for probabilistic time-series imputation and conditional modeling~\cite{tashiro2021csdi}, while Diffusion-TS extended diffusion models to general time-series generation~\cite{yuan2024diffusionts}. These approaches model temporal distributions rather than relying on pointwise predictions.

Recent work has extended this paradigm toward richer conditioning and sparse
contexts. TimeWeaver~\cite{narasimhan2024timeweaver} conditions diffusion-based generation on heterogeneous
static and time-varying metadata while preserving their temporal alignment
with the generated trajectory. CENTS~\cite{fuest2025cents} targets
rare and unseen context combinations through a context encoder and a
context-dependent normalizer that predicts target statistics from
context. Complementarily, TimeDP~\cite{huang2025timedp} enables multi-domain time-series generation by adapting to new domains using prompts extracted from a small amount of target data. These approaches respectively address
heterogeneous conditions, generalization to sparse contexts, and adaptation
from limited observations.

Our novel approach is to bring these directions together specifically for RAN energy what-if generation. In fact, we combine temporally aligned conditioning with configuration-dependent normalization, then introduce a prediction-based decomposition for sparse configuration and operating-regime coverage.

\section{RAN Optimization Use Case}
\label{sec:problem}

RAN energy-saving mechanisms operate at different spatial and temporal granularities, including symbol-level shutdown, antenna or MIMO muting, and carrier shutdown. A common control strategy relies on hysteresis: resources (e.g. carriers or cells) are deactivated when traffic load falls below $\rho_{\min}$ and reactivated when it exceeds $\rho_{\max}$, while sufficient resources remain active to preserve coverage~\cite{maggi2023energy}. Selecting these predefined thresholds requires assessing how a candidate policy behaves across varying traffic and operating conditions. In practice, historical measurements cover only a limited range of radio configurations and operating conditions, which together define the context considered in this work. Some contexts therefore have little or no training data, while particular operating regimes may be missing even for otherwise well-represented radio configurations. This motivates the use of conditional generative AI methods for what-if analysis, able to generate plausible power consumption trajectories for a given radio site and configuration under requested operating scenarios, including hypothetical traffic patterns not yet observed in the live network.

\subsection{Conditional Generation for What-If Analysis}

For a given radio site and configuration $j$, we define a power consumption \emph{trajectory} as a sequence of $L$ consecutive power consumption measurements, $\mathbf{y}^{(i)}_j=(y^{(i)}_{j,1},\ldots,y^{(i)}_{j,L})$. This trajectory is influenced by a context $\mathbf{c}^{(i)}_j=(\mathbf{m}_j,\mathbf{x}^{(i)}_j)$, where $\mathbf{m}_j$ denotes the \emph{static configuration}, including site identity and hardware characteristics, and $\mathbf{x}^{(i)}_j =
(\mathbf{x}^{(i)}_{j,1},\ldots,\mathbf{x}^{(i)}_{j,L})$ denotes the time-varying \emph{operating context} for trajectory $i$ under configuration $j$, including traffic load (e.g., PRB load) and energy-saving activity. We assume that both the static configuration and the time-varying operating conditions are specified by the requested what-if scenario and used as conditioning inputs.

The objective of conditional generation is to learn the distribution of power consumption trajectories given their context. A synthetic trajectory is then drawn from the conditional distribution as

\begin{equation}
    \tilde{\mathbf{y}}^{(i)}_j
    \sim
    p_\theta\!\left(
    \cdot \mid \mathbf{c}^{(i)}_j
    \right),
    \label{eq:conditional_generation}
\end{equation}
where $\theta$ denotes the learned model parameters.

Given a hypothetical context, repeated sampling produces multiple temporally coherent power consumption trajectories, representing plausible outcomes of the same what-if scenario. This differs from pointwise prediction methods, which produce predictions at individual time steps rather than sampling from the joint conditional distribution of a complete trajectory.

During training, we distinguish three levels of exposure according to the amount of data available for a given context.  A context may be \emph{well-seen}, or \emph{well-represented}, in the training set, with multiple training trajectories; \emph{rarely seen}, with only a single training trajectory; or \emph{unseen}, with no training trajectory.
Independently, at inference, the trajectory generation for a given context is qualified as \emph{zero-shot} when no target power observation is available and \emph{few-shot} when only a limited number $K>0$ are available.

\subsection{Datasets and Context Sparsity}

\begin{table}[b]
\refstepcounter{table}
\label{tab:dataset_description}
{\footnotesize Table~\thetable. Datasets used in the evaluation.\par}
\vspace{5pt}
\scriptsize
\centering
\begin{tabular}{|p{1.1cm}|p{0.6cm}|p{0.4cm}|p{0.9cm}|p{0.6cm}|p{1.2cm}|p{0.8cm}|}
\hline
\textbf{Dataset} &
\textbf{BSIDs} &
\textbf{Days} &
\textbf{Sampling} &
\textbf{Points} &
\textbf{Trajectories} &
\textbf{Groups} \\
\hline
Private & 9 & 14 & 15 min & 42,031 & 823 & 65 \\
Public~\cite{huawei2023energy} & 923 & 8 & 60 min & 92,629 & N/A & 923 \\
\hline
\end{tabular}
\end{table}

\begin{table}[b]
\refstepcounter{table}
\label{tab:private_variables}
{\footnotesize
Table~\thetable. Features in the datasets
{\scriptsize ($^1$specific to private RAN, $^2$specific to
public~\cite{huawei2023energy}).}\par}
\vspace{5pt}
\scriptsize
\centering
\begin{tabular}{|p{2cm}|p{4.4cm}|p{1.2cm}|}
\hline
\textbf{Role} & \textbf{Variables} & \textbf{Type} \\
\hline
Site & base station ID & Categorical \\
Hardware configuration (static) & Radio module, product, 5G band, number of cells, bandwidth$^2$, tx\_power$^2$ & Categorical \\
Operating context (dynamic) & Traffic load, cell energy-saving duty cycle$^1$, MIMO muting duty-cycle$^1$, power saving mode$^2$ & Continuous \\
Time & Minutes, hour, day of week & Categorical \\
Target  & Average power consumption (Watts) & Continuous \\
\hline
\end{tabular}
\end{table}

We use two complementary 5G network datasets described in Table~\ref{tab:dataset_description}, whose features are summarized in Table~\ref{tab:private_variables}. The two datasets differ in structure and context diversity, providing complementary settings for the conditioning setup and the challenges considered in the conditional-generation task.  The public ITU dataset, with its large number of base stations, is particularly well suited for evaluating the global power predictor and the residual level and scale correction mechanisms of the prediction-based framework. In contrast, the longer contiguous trajectories available in the private dataset enable a complete evaluation of trajectory generation and a direct comparison between the direct and prediction-based generative approaches.

All observations sharing the same site identity and static hardware configuration, i.e., the same static configuration $x^{(i)}_j$, define a \emph{group} of trajectories.  In the private dataset, a group is identified by the tuple (site, radio module, product, 5G band, number of cells). 
Using the base station ID (BSID) alone as the grouping variable would yield only 9 groups, providing insufficient diversity in static configurations for the conditional-generation study. We therefore define groups at a finer aggregation level by combining the site identifier with the radio hardware configuration described in Table~\ref{tab:private_variables}.
In the public dataset~\cite{huawei2023energy}, the group corresponds directly to a BSID.

As detailed in Table~\ref{tab:dataset_description}, the private dataset comprises operational measurements collected from a commercial 5G RAN, with 42,031 samples recorded every 15 minutes across 65 groups and 9 base stations. From contiguous time series, 823 power consumption trajectories of length $L=48$, corresponding to 12-hour windows are extracted. The public dataset contains 92,629 hourly measurements from 923 base stations, each defining one group. The larger population of the public dataset makes it particularly suitable for studying transfer and few-shot correction for rarely seen and unseen base stations, although its temporal coverage does not provide sufficiently long contiguous sequences for trajectory-generation experiments. Conversely, longer contiguous trajectories available in the private dataset make it better suited for studying power-consumption trajectory generation.

\begin{figure}[!t]
\centering
\includegraphics[width=0.9\columnwidth]{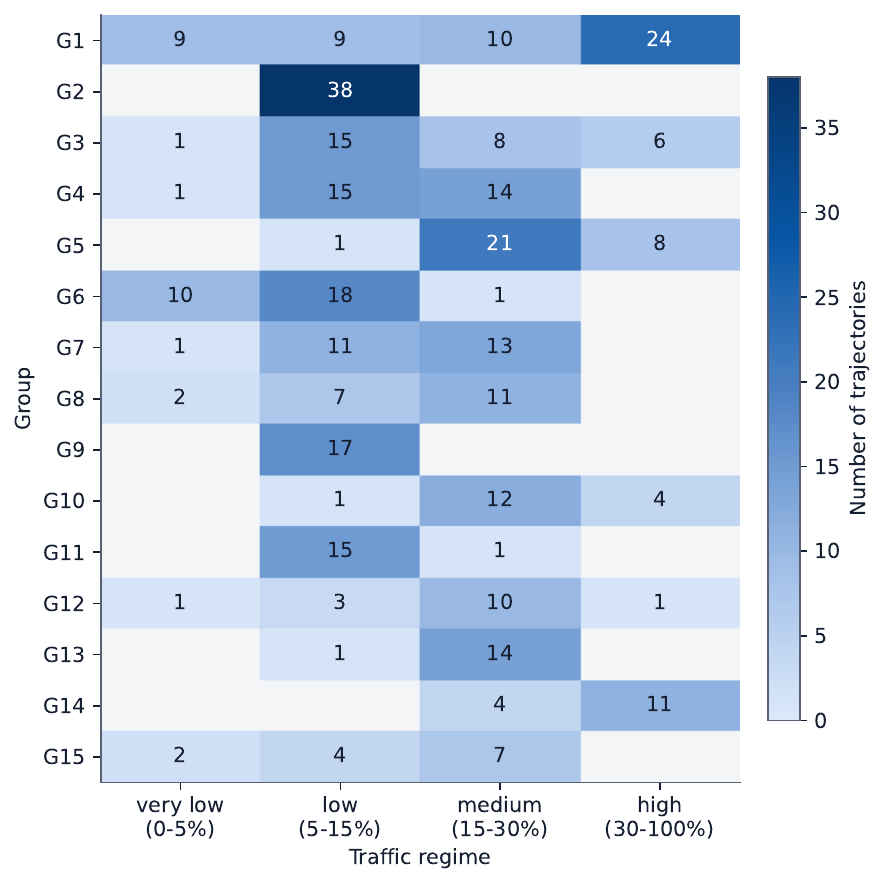}
\caption{Training-trajectory coverage across traffic regimes for the 15 most represented groups in the private dataset. Each cell reports the number of trajectories in the corresponding regime.}
\label{fig:context_sparsity_heatmap}
\end{figure}

\begin{figure}[t]
\centering
\includegraphics[width=\columnwidth]{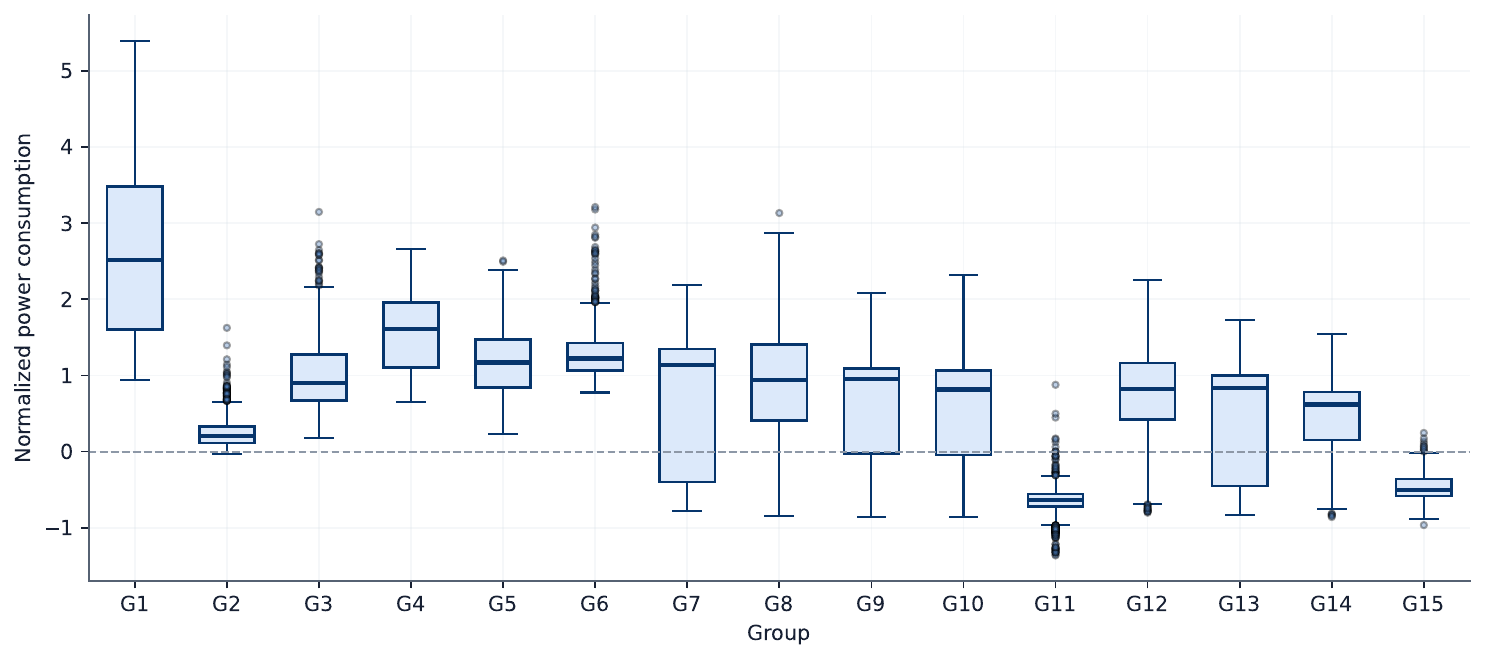}
\caption{Globally normalized power consumption across the 15 most represented groups in the private dataset. Groups retain substantially different levels and dispersions.}
\label{fig:power_scale_by_group}
\end{figure}

The data exhibit two additional challenges. First, training coverage is highly uneven across groups and operating regimes. Fig.~\ref{fig:context_sparsity_heatmap} illustrates this sparsity in the private dataset. Even among well-seen groups, training trajectories are unevenly distributed across traffic-load regimes. Hence, conditional generation must capture the temporal relationship between operating conditions and power consumption while remaining robust to both poorly represented groups and operating regimes absent for a given group. 
Second, power consumption varies substantially across groups. Fig.~\ref{fig:power_scale_by_group} shows marked group-specific differences in consumption level and dispersion in the private dataset. A conditional generator must therefore learn not only the temporal variations, but also the mean level and variability specific to each context.
This heterogeneity motivates explicitly accounting for group-dependent consumption level and scale in the proposed models.

\section{Proposed Approaches}
\label{sec:approach}

In this paper, we propose two complementary formulations tailored for RAN energy what-if generation. The first directly generates a power consumption trajectory from the requested context. To better handle groups with little or no training data, the second decomposes generation into global power prediction, residual level and scale correction, and stochastic temporal residual generation. The temporal component can reuse the conditional generator introduced in the first formulation, applied to standardized residual trajectories.

\subsection{Context-Normalized TimeWeaver}
\label{subsec:cn_timeweaver}

\begin{figure}[t]
    \centering
    \includegraphics[width=0.8\columnwidth]{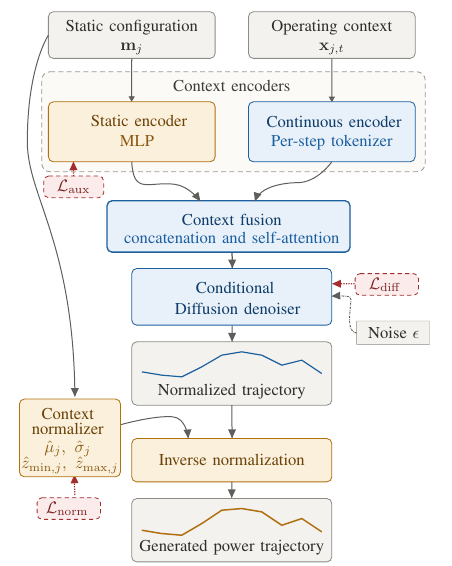}
    \caption{Context-Normalized TimeWeaver architecture. Orange components are inspired by CENTS, blue components follow the TimeWeaver backbone, and red dashed boxes indicate training losses.}
    \label{fig:cn_timeweaver}
\end{figure}

Our direct approach combines the temporally aligned conditioning of TimeWeaver~\cite{narasimhan2024timeweaver} with context encoding and configuration-dependent normalization inspired by CENTS~\cite{fuest2025cents}. 
This resulting model is denoted Context-Normalized TimeWeaver (CN-TimeWeaver), and its architecture is shown in
Fig.~\ref{fig:cn_timeweaver}.

Each categorical field of the static configuration $\mathbf m_j$ is embedded
separately and compressed by a Multi-Layer Perceptron (MLP) with a
low-dimensional bottleneck. This representation is broadcast over the
trajectory, while the operating context $\mathbf x_{j,t}$ is encoded at every
time step $t$. The two representations are fused to condition the diffusion
denoiser, preserving their temporal alignment with the generated trajectory.
Following TimeWeaver, we use the CSDI backbone~\cite{tashiro2021csdi},
originally designed for time-series imputation and adapted by TimeWeaver to
metadata-conditioned generation.

Power consumption level and scale vary across groups, motivating
configuration-dependent target normalization. For each training group, we compute the power mean $\mu_j$ and standard
deviation $\sigma_j$, together with the minimum $z_{\min,j}$ and maximum $z_{\max,j}$ after standardization. A parametric normalizer predicts these four statistics from the static configuration. The power is then mapped to
the normalized domain used by the denoiser as

\begin{equation}
    y'_{j,t}
    =
    \frac{1}{\hat z_{\max,j}-\hat z_{\min,j}+\delta}
    \left(
        \frac{y_{j,t}-\hat\mu_j}{\hat\sigma_j+\delta}
        -\hat z_{\min,j}
    \right),
    \label{eq:cn_normalization}
\end{equation}

where $\delta$ prevents division by zero. The same
predicted statistics invert the transformation after generation. Hence, inference requires only the requested context and no target power observation.

The model is trained jointly using

\begin{equation}
    \mathcal L
    =
    \mathcal L_{\mathrm{diff}}
    +
    \lambda_{\mathrm{aux}}\mathcal L_{\mathrm{aux}}
    +
    \lambda_{\mathrm{norm}}\mathcal L_{\mathrm{norm}},
    \label{eq:cn_tw_loss}
\end{equation}

where $\mathcal L_{\mathrm{diff}}$ is the DDPM (Denoising Diffusion Probabilistic Models) loss~\cite{ho2020ddpm}, $\mathcal L_{\mathrm{aux}}$ reconstructs the static configuration from its compressed representation, and $\mathcal L_{\mathrm{norm}}$ penalizes errors on the predicted normalization statistics. The weights $\lambda_{\mathrm{aux}}$ and $\lambda_{\mathrm{norm}}$ control the relative contributions of the auxiliary and normalization losses. CN-TimeWeaver therefore learns power consumption level, scale, and temporal variation from the requested configuration and operating conditions.

\subsection{Prediction-Based Residual Correction}
\label{subsec:prediction_based}

The second approach separates power generation into three successive steps. 
For group $j$ at time $t$, with the context $\mathbf c_{j,t}=(\mathbf m_j,\mathbf x_{j,t})$, we decompose the power consumption as the sum of three components:
\begin{equation}
    y_{j,t}
    =
    f(\mathbf c_{j,t})
    +
    b_j
    +
    s_j z_{j,t},
    \label{eq:residual_decomposition_paper}
\end{equation}

where $f(\mathbf c_{j,t})$ is the global prediction, $b_j$ is the residual level, $s_j$ is the residual scale, and $z_{j,t}$ is the  standardized temporal residual.

First, a global MLP-based power predictor $f$ estimates consumption from the context.
Inspired by~\cite{chen2025energyfingerprint}, the predictor
combines the static configuration, standardized operating variables, and
calendar variables with a learned embedding of the base station identity. This embedding is randomly masked during training so that the model can also operate when the target identity was not observed during training.
The resulting representation is processed by two hidden layers of widths 128
and 64. 

Second, the residual level and scale are estimated separately. When $K>0$
target observations are available, the residual level is estimated as the mean prediction residual over these observations: 

\begin{equation}
    \hat b_j
    =
    \frac{1}{K}
    \sum_{t \in \mathcal S_j}
    \left[
        \ y_{j,t}-f(\mathbf c_{j,t})
    \ \right],
    \label{eq:residual_level}
\end{equation}
where $\mathcal S_j$ denotes the set of $K$ target observations. In the
zero-shot setting (i.e., $K=0$), we set $\hat b_j=0$. The residual scale $s_j$ is estimated without
target power observations. For each group $j$ used to fit the scale model, we compute the standard deviation of its prediction residuals after removing
their mean. A Ridge regression then learns to predict the logarithm of this scale from static hardware metadata. For a target group, this model directly provides $\hat s_j$ from its metadata.

Finally, the standardized temporal residual $z_{j,t}$ is generated using
either an AR(3) process with Student's-$t$ innovations or TimeWeaver trained separately for each group on standardized prediction residual trajectories, using the same temporally aligned operating context as conditioning. We denote the resulting corrected models C-AR(3) and C-TimeWeaver, where C stands for \emph{Corrected}. Both are fitted to standardized prediction residuals rather than power trajectories.

Generation then follows Eq.~\eqref{eq:residual_decomposition_paper} using
$\hat b_j$, $\hat s_j$, and a sampled residual $\tilde z_{j,t}$. The global
predictor and residual generator are trained once and remain unchanged when
target observations are added. Few-shot adaptation therefore updates only
$\hat b_j$, while $\hat s_j$ remains determined by metadata.
 
\section{Numerical Results}
\label{sec:results}

We evaluate the proposed approaches across training exposure, inference support, and unseen operating regimes.

\subsection{Experimental Protocol}
\label{subsec:protocol}

Trajectory generation is evaluated on the private dataset, while both datasets
are used to evaluate the global predictor $f$ and the residual correction terms
$\hat b_j$ and $\hat s_j$. Private trajectories have length $L=48$ samples (corresponding to 12\,h), using a stride of 24 samples.

We use the same 41 groups for the training-exposure and inference-support experiments. For each group, the $4$ most recent trajectories are kept for evaluation. Training uses only earlier trajectories, and any trajectory that
overlaps these four evaluation trajectories is removed. Target observations
used for few-shot correction always precede these trajectories and never
overlap with them.
In the well-seen setting, all groups with at least $5$ available earlier training trajectories are considered, and one generator is trained using all available earlier trajectories. In the rarely seen setting, only one earlier trajectory from the evaluated group is kept for training. In the unseen setting, no trajectory from that group is used in training. For the latter two settings, we train a separate generator for each evaluated group, modifying only the training exposure of the target group while retaining all eligible training data from the other groups.

For C-TimeWeaver, the global power predictor $f$ is always trained without power measurements from the evaluated group. Only the amount of data available to
the residual generator changes across the three settings.

We evaluate point accuracy with Mean Absolute Error (MAE), temporal similarity
with Dynamic Time Warping (DTW)~\cite{berndt1994dynamic}, and probabilistic
accuracy with the Continuous Ranked Probability Score
(CRPS)~\cite{gneiting2007scoring}. Lower values are better for all three metrics. For each group, $4$ real trajectories are retained for evaluation. For each generation seed and evaluation trajectory, $8$ synthetic trajectories are generated. MAE and DTW are computed for each generated trajectory and then averaged over the $8$ samples, whereas CRPS is computed jointly using the $8$ generated trajectories. The resulting metrics are then averaged over the $4$ evaluation trajectories and over three generation seeds. Finally, results are summarized across the 41 groups using the median and Median Absolute Deviation (MAD).
The remaining training data are split into training and validation subsets, with validation used only for model selection. The neural models are implemented in PyTorch and trained on a NVIDIA GTX Titan X.

Across the 41 groups used for evaluation, the mean power consumption over the evaluation trajectories is approximately $280$\,W. This provides a useful reference for the reported MAEs, which range from a few watts in the best-performing settings to about $46$\,W in the most challenging unseen case.

In terms of model size, TimeWeaver and its residual counterpart each contain $0.66$M trainable parameters, while CN-TimeWeaver contains $0.69$M. The global power predictor contains $31.7$k parameters, and the residual scale Ridge model and AR(3) require only 34 coefficients and $5$ scalars, respectively.  

\begin{table}[t]
\refstepcounter{table}
{\footnotesize Table~\thetable. Conditional generation as the amount of training data from the evaluated group
decreases. Values are median $\pm$ MAD across groups after
averaging over three generation seeds within each group. Bold marks the best
result within each exposure regime. Every row is zero-shot.\par}
\vspace{5pt}
\label{tab:conditional_exposure}
\centering
\scriptsize
\setlength{\tabcolsep}{4.3pt}
\begin{tabular}{llccc}
\toprule
Exposure & Method
& MAE (W) $\downarrow$
& DTW $\downarrow$
& CRPS (W) $\downarrow$ \\
\midrule

\multirow{3}{*}{Well-seen}
& TimeWeaver
& $10.8\!\pm\!3.9$
& $0.513\!\pm\!0.170$
& $6.2\!\pm\!2.0$ \\

& CN-TimeWeaver
& $4.6\!\pm\!2.3$
& $\mathbf{0.273\!\pm\!0.132}$
& $\mathbf{3.3\!\pm\!1.6}$ \\

& C-TimeWeaver
& $\mathbf{4.5\!\pm\!2.5}$
& $0.339\!\pm\!0.180$
& $3.6\!\pm\!1.8$ \\

\midrule

\multirow{3}{*}{Rarely seen}
& TimeWeaver
& $16.7\!\pm\!5.0$
& $0.525\!\pm\!0.164$
& $\mathbf{8.0\!\pm\!2.3}$ \\

& CN-TimeWeaver
& $15.7\!\pm\!11.6$
& $\mathbf{0.415\!\pm\!0.297}$
& $11.3\!\pm\!8.3$ \\

& C-TimeWeaver
& $\mathbf{14.3\!\pm\!10.3}$
& $0.594\!\pm\!0.368$
& $12.7\!\pm\!10.0$ \\

\midrule

\multirow{3}{*}{Unseen}
& TimeWeaver
& $17.1\!\pm\!6.0$
& $\mathbf{0.543\!\pm\!0.185}$
& $\mathbf{8.7\!\pm\!3.4}$ \\

& CN-TimeWeaver
& $45.9\!\pm\!27.2$
& $1.205\!\pm\!0.590$
& $44.2\!\pm\!30.4$ \\

& C-TimeWeaver
& $\mathbf{14.1\!\pm\!9.8}$
& $0.617\!\pm\!0.363$
& $12.8\!\pm\!9.3$ \\

\bottomrule
\end{tabular}
\end{table}

\begin{figure}[b]
\centering
\includegraphics[width=0.9\columnwidth]{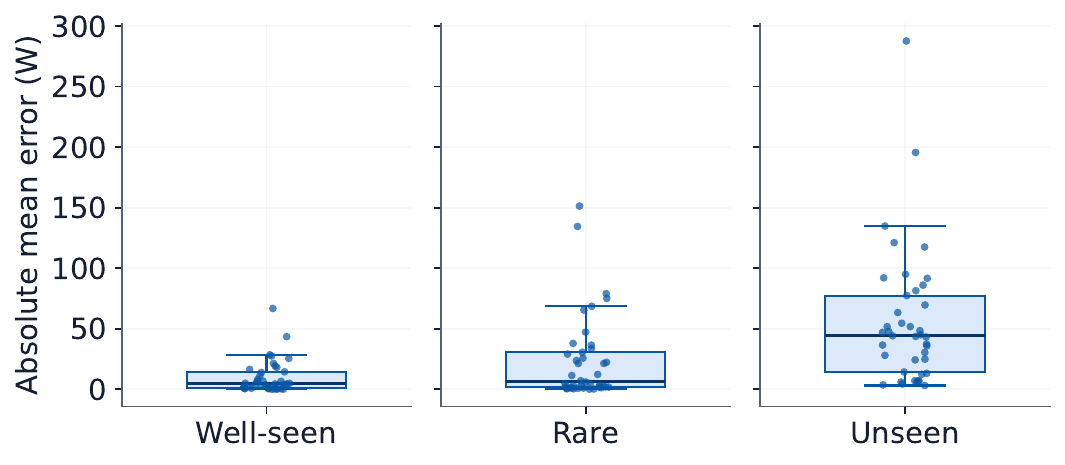}
\caption{Absolute error of the configuration-dependent normalization mean predicted by CN-TimeWeaver as the amount of training data from the evaluated group decreases. Points correspond to evaluated groups and
boxes summarize their distribution.
}
\label{fig:cn_tw_normalizer_mu_error}
\end{figure}

\subsection{Conditional Generation with Decreasing Training Data}
\label{subsec:conditional_results}

We first examine how the three main conditional generators behave as the amount of
training data from the evaluated group decreases. 
In this comparative study, TimeWeaver serves as the direct baseline, CN-TimeWeaver incorporates the CENTS-inspired encoder and normalizer, and C-TimeWeaver represents the prediction-based residual correction method.
Table~\ref{tab:conditional_exposure} shows two distinct behaviors. When the
evaluated group is well represented, both proposed methods substantially
improve over TimeWeaver. CN-TimeWeaver gives the best DTW and CRPS, while
C-TimeWeaver gives the lowest MAE. With only one training trajectory from the
evaluated group (i.e. rarely seen), CN-TimeWeaver retains a DTW improvement but loses
probabilistic accuracy. When no trajectory from the group is used for training (i.e. unseen),
CN-TimeWeaver degrades sharply on all three metrics. C-TimeWeaver remains much
more stable as training exposure decreases, although without target
observations at inference it does not outperform TimeWeaver on DTW or CRPS.
C-AR(3) is not included in this comparison because its simple residual dynamics
are not competitive in the well-seen regime. We instead evaluate it as a
lightweight residual baseline in the inference-support study.

Fig.~\ref{fig:cn_tw_normalizer_mu_error} helps to explain the degradation of
CN-TimeWeaver. Errors in the predicted normalization mean are small for most well-seen groups and increase as training exposure decreases. For some unseen
groups, the generated trajectories remain shifted from the real consumption
level despite preserving plausible temporal variation. This motivates
estimating the consumption level separately from residual generation.

\begin{table}[!t]
\refstepcounter{table}
{\footnotesize Table~\thetable. Prediction-based generation for unseen groups
as a function of inference support on the private dataset. Values are median
$\pm$ MAD across groups after averaging over three generation seeds. 
Standard TimeWeaver does not use target power observations at inference (i.e., $K >0$).
Bold marks the best residual method at each support size.\par}
\vspace{5pt}
\label{tab:prediction_based_generation}
\centering
\scriptsize
\renewcommand{\arraystretch}{1.12}

\begin{tabular*}{\columnwidth}{
    @{\extracolsep{\fill}}
    l
    l
    c
    c
    c
    @{}
}
\toprule
$K$ / Time & Method
& MAE (W) $\downarrow$
& DTW $\downarrow$
& CRPS (W) $\downarrow$ \\
\midrule

\multirow{3}{*}{$0$}

& TimeWeaver
& $17.1\!\pm\!6.0$
& $0.543\!\pm\!0.185$
& $8.7\!\pm\!3.4$ \\
& C-AR(3)
& $14.7\!\pm\!10.9$
& $0.654\!\pm\!0.354$
& $\mathbf{12.7\!\pm\!9.7}$ \\

&
C-TimeWeaver
& $\mathbf{14.1\!\pm\!9.8}$
& $\mathbf{0.617\!\pm\!0.363}$
& $12.8\!\pm\!9.3$ \\[1.5pt]

\cdashline{1-5}[0.4pt/1.3pt]
\noalign{\vskip 3pt}
\multirow{2}{*}{$1$ / $15$min} 
& TimeWeaver
& -
& -
& - \\
& C-AR(3)
& $\mathbf{8.3\!\pm\!4.8}$
& $0.462\!\pm\!0.259$
& $\mathbf{6.8\!\pm\!4.2}$ \\

&
C-TimeWeaver
& $8.7\!\pm\!5.6$
& $\mathbf{0.435\!\pm\!0.251}$
& $6.9\!\pm\!4.5$ \\[1.5pt]

\cdashline{1-5}[0.4pt/1.3pt]
\noalign{\vskip 3pt}

\multirow{2}{*}{$8$ / $2$h}
& TimeWeaver
& -
& -
& - \\
& C-AR(3)
& $7.7\!\pm\!4.8$
& $0.398\!\pm\!0.221$
& $6.3\!\pm\!4.0$ \\

&
C-TimeWeaver
& $\mathbf{7.2\!\pm\!4.7}$
& $\mathbf{0.395\!\pm\!0.211}$
& $\mathbf{5.6\!\pm\!3.8}$ \\[1.5pt]

\cdashline{1-5}[0.4pt/1.3pt]
\noalign{\vskip 3pt}

\multirow{2}{*}{$48$ / $12$h}
& TimeWeaver
& -
& -
& - \\
& C-AR(3)
& $6.9\!\pm\!4.0$
& $0.409\!\pm\!0.213$
& $5.3\!\pm\!3.1$ \\

&
C-TimeWeaver
& $\mathbf{5.8\!\pm\!3.5}$
& $\mathbf{0.396\!\pm\!0.195}$
& $\mathbf{5.1\!\pm\!3.1}$ \\

\bottomrule
\end{tabular*}
\end{table}

\subsection{Prediction-Based Correction and Ablation Study}
\label{subsec:correction_results}

In this section, we analyze inference-time support separately from training exposure, which is studied
independently in Table~\ref{tab:conditional_exposure}.  

We first examine the impact of both correction terms $b_j$ and $s_j$  independently of trajectory
generation. Across both datasets, hardware metadata is more informative about
residual scale than residual level, with an out-of-fold $R^2$ of about $0.6$
for log-scale prediction, while level prediction remains weak. Conversely,
target observations provide valuable information about the residual level. On unseen
base stations in the public dataset, even one target observation yields a
marked improvement, showing that the same decomposition is also effective
beyond the private dataset.

\begin{figure}[t]
    \centering
    \includegraphics[width=\columnwidth]{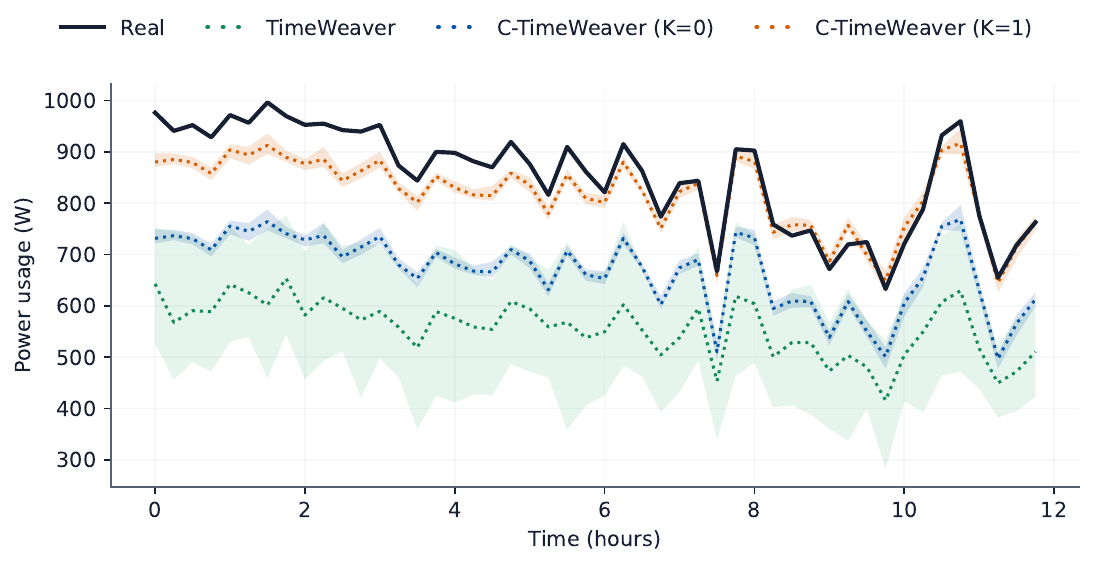}
    \caption{Direct and prediction-based generation without target observations ($K=0$) and with residual-level correction using one target observation ($K=1$), for a challenging unseen group.}
    \label{fig:unseen_qual}
\end{figure}

We next evaluate the two complete prediction-based generators on unseen groups of
the private dataset. Table~\ref{tab:prediction_based_generation} varies the
number $K$ of target observations available at inference while keeping the
global predictor and residual generator fixed. At the 15-minute sampling rate,
the three non-zero support sizes correspond to 15 minutes, 2 hours, and
12 hours of target observations. 
At $K=0$, the prediction-based approach mainly improves consumption-level
accuracy and does not consistently improve all trajectory metrics. However, once target observations are available, both corrected generators improve substantially. Table~\ref{tab:prediction_based_generation} shows that, with only one target observation (i.e., $K=1$), the median MAE is reduced by about $50\%$ compared with TimeWeaver, from $17.1$\,W to $8.3$\,W for C-AR(3) and $8.7$\,W for C-TimeWeaver. 
Since few-shot adaptation updates only $\hat{b}_j$, these gains highlight the
importance of residual-level correction. C-TimeWeaver achieves the best
temporal agreement as support increases, while C-AR(3) remains a competitive
lightweight alternative.
Fig.~\ref{fig:unseen_qual} illustrates the correction mechanism on a challenging unseen group. At $K=0$, both direct and corrected generation remain below the
real consumption level, although C-TimeWeaver reduces the shift. With only one target observation, updating $\hat b_j$ moves the corrected trajectory close
to the observed consumption level without retraining either the predictor or the residual generator.
Similar results are observed for rarely seen groups, with an MAE reduction exceeding $50\%$ at $K=1$.

Finally, we ablate the components of
Eq.~\eqref{eq:residual_decomposition_paper} on unseen groups at $K=48$ to assess their impact on trajectory generation.
Table~\ref{tab:residual_ablation} reveals a clear hierarchy between the
components. Residual-level correction provides most of the gain. Without it, adding a
stochastic residual has little effect on MAE and improves CRPS at the expense
of DTW. Once the level is corrected, residual generation further improves
CRPS, and C-TimeWeaver also reduces MAE, while the deterministic
$f+\hat b$ reconstruction retains the lowest DTW. The relatively small gap
between AR(3) and TimeWeaver compared with the gain from $\hat b$ indicates
that accurate level estimation is the dominant component.

\subsection{What-If Generation Under an Unseen Operating Regime}
\label{subsec:whatif_operating_regime}

Finally, we evaluate a what-if setting where the equipment configuration is represented during training but a specific operating regime is not. For $8$ groups, high-traffic trajectories with load above $50\%$ are removed from training. Lower-traffic trajectories from the same group remain available, as do high-traffic trajectories from other groups. Evaluation is restricted to the held-out high-traffic trajectories using their measured operating context.
The experiment therefore tests whether a generator can reproduce power under a traffic regime that is globally represented but unseen for the requested equipment configuration. Table~\ref{tab:unseen_operating_regime} reports the generation performance on these held-out high-traffic trajectories.

\begin{table}[t]
\refstepcounter{table}
{\footnotesize Table~\thetable. Ablation of the prediction-based decomposition
on unseen groups at $K=48$. $z_{\mathrm{AR(3)}}$ and $z_{\mathrm{TW}}$
denote standardized residual trajectories generated by the AR(3) and
TimeWeaver residual models, respectively. Values are median $\pm$ MAD across
groups. Bold marks the best result.\par}
\vspace{5pt}
\label{tab:residual_ablation}
\centering
\scriptsize
\renewcommand{\arraystretch}{1.12}

\begin{tabular*}{\columnwidth}{
    @{\hspace{0.7em}\extracolsep{\fill}}
    c
    l
    c
    c
    c
    @{}
}
\toprule
& Reconstruction
& MAE (W) $\downarrow$
& DTW $\downarrow$
& CRPS (W) $\downarrow$ \\
\midrule

\multirow{3}{*}{\rotatebox[origin=c]{90}{Without $\hat b$}}
& $f$
& $14.6\!\pm\!11.1$
& $0.582\!\pm\!0.382$
& $14.6\!\pm\!11.1$ \\

& $f+\hat s z_{\mathrm{AR(3)}}$
& $14.7\!\pm\!10.9$
& $0.654\!\pm\!0.354$
& $12.7\!\pm\!9.7$ \\

& $f+\hat s z_{\mathrm{TW}}$
& $14.1\!\pm\!9.8$
& $0.617\!\pm\!0.363$
& $12.8\!\pm\!9.3$ \\[1.5pt]

\midrule

\multirow{3}{*}{\rotatebox[origin=c]{90}{With $\hat b$}}
& $f+\hat b$
& $6.7\!\pm\!4.0$
& $\mathbf{0.331\!\pm\!0.214}$
& $6.7\!\pm\!4.0$ \\

& $f+\hat b+\hat s z_{\mathrm{AR(3)}}$
& $6.9\!\pm\!4.0$
& $0.409\!\pm\!0.213$
& $5.3\!\pm\!3.1$ \\

& $f+\hat b+\hat s z_{\mathrm{TW}}$
& $\mathbf{5.8\!\pm\!3.5}$
& $0.396\!\pm\!0.195$
& $\mathbf{5.1\!\pm\!3.1}$ \\

\bottomrule
\end{tabular*}
\end{table}

C-TimeWeaver performs best on all three metrics. The distinction between
C-TimeWeaver and CN-TimeWeaver is particularly relevant in this experiment.
As illustrated in Fig.~\ref{fig:power_scale_by_group}, consumption level and
dispersion vary substantially across groups. The global predictor $f$ used by
C-TimeWeaver receives the time-varying operating context, including the traffic load, and can therefore adapt its power estimate to the requested traffic regime.
In CN-TimeWeaver, the diffusion denoiser also receives the operating context,
but the normalization statistics are predicted only from the static
configuration. The same group-level normalization is therefore used across
different traffic regimes. This helps explain the stronger generalization of
C-TimeWeaver when the operating regime is unseen for the requested group.

\subsection{Practical Deployment Guidance}
\label{subsec:discussion}

\begin{figure}[!t]
    \centering
    \includegraphics[width=\columnwidth]{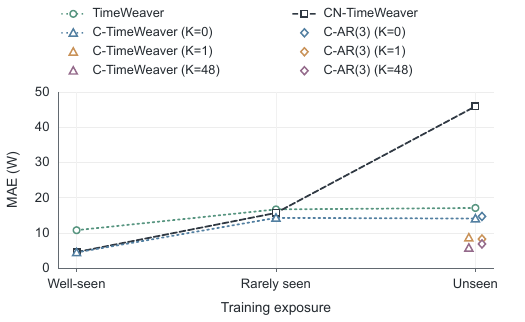}
    \caption{Median MAE across training-exposure regimes and inference support. Few-shot results ($K>0$) are reported for unseen groups. C-AR(3) is shown only in the unseen setting.}
    \label{fig:mae_training_exposure}
\end{figure}

Fig.~\ref{fig:mae_training_exposure} summarizes the main trade-offs for practical deployment. For well-seen configurations, CN-TimeWeaver and C-TimeWeaver provide similar MAE improvements over TimeWeaver, with CN-TimeWeaver also achieving the best DTW and CRPS. As training exposure decreases, however, CN-TimeWeaver becomes less reliable, while zero-shot C-TimeWeaver remains more stable in terms of MAE. The choice of method should therefore account for both training coverage and the metric relevant to the intended what-if analysis.

When target observations are available, the prediction-based framework provides an additional adaptation mechanism. For unseen groups, using only one target observation reduces the median MAE by about $50\%$ relative to the zero-shot TimeWeaver baseline, while additional observations further improve accuracy. In our setup, standard TimeWeaver does not use target power observations at inference without updating the model, whereas the prediction-based approach only requires recomputing the residual-level correction $\hat b_j$, without retraining the predictor or residual generator. As inference support increases (i.e., $K>0$), C-TimeWeaver generally provides the best trajectory-generation accuracy, while C-AR(3) remains a competitive lightweight alternative.

Computational cost further differentiates the two residual generators. In our implementation and setup using an NVIDIA RTX 3800 GPU, generating $8$ trajectories of length $48$ with diffusion-based generation and $350$ reverse steps takes about $16$\,s, compared with about $5$\,ms for global prediction, scale estimation, AR(3) generation, and few-shot correction. This difference is mainly due to iterative diffusion sampling. C-AR(3) therefore offers a low-cost option when latency or large-scale generation is a primary constraint, whereas C-TimeWeaver trades additional computation for improved generation accuracy when sufficient inference support is available. These runtime figures are hardware- and implementation-dependent and apply independently of the training-exposure regime.

\section{Conclusion}
\label{sec:conclusion}

We studied conditional generation of RAN power consumption trajectories under limited training coverage of both radio configurations and operating regimes. CN-TimeWeaver substantially improves generation for well-seen groups, but its configuration-dependent normalization becomes less reliable as training exposure decreases. To address rarely seen and unseen groups, we introduced a prediction-based framework that separates global power prediction,
residual level and scale correction, and temporal residual generation. Results show that hardware metadata provides transferable information about residual scale, while limited target observations used at inference correct residual level and substantially improve generation without retraining the predictor or generator. With only one single target power observation, the prediction-based methods reduce the median MAE by about $50\%$ compared with TimeWeaver for rarely and unseen contexts.
Our ablation study confirms that residual-level correction provides most of the gain, while the prediction-based approach also generalizes better to a high-traffic regime unseen for an otherwise represented group. Future work will investigate improved zero-shot correction and probabilistic calibration, as well as more complex what-if scenarios for RAN energy-saving policies.

\begin{table}[!t]
\refstepcounter{table}
{\footnotesize Table~\thetable. Generation under an unseen high-traffic
operating regime for $8$ represented groups. Values are median $\pm$ MAD
across groups after averaging over three generation seeds within each group.
Bold marks the best result.\par}
\vspace{5pt}
\label{tab:unseen_operating_regime}
\centering
\scriptsize
\renewcommand{\arraystretch}{1.12}

\begin{tabular*}{\columnwidth}{
    @{\extracolsep{\fill}}
    l
    c
    c
    c
    @{}
}
\toprule
Method
& MAE (W) $\downarrow$
& DTW $\downarrow$
& CRPS (W) $\downarrow$ \\
\midrule

TimeWeaver
& $15.5\!\pm\!5.8$
& $0.618\!\pm\!0.258$
& $10.4\!\pm\!4.0$ \\

CN-TimeWeaver
& $12.1\!\pm\!2.8$
& $0.596\!\pm\!0.200$
& $9.1\!\pm\!1.6$ \\

C-TimeWeaver
& $\mathbf{9.2\!\pm\!2.6}$
& $\mathbf{0.443\!\pm\!0.134}$
& $\mathbf{7.5\!\pm\!2.1}$ \\

\bottomrule
\end{tabular*}
\end{table}

\balance
\bibliographystyle{IEEEtran}
\bibliography{biblio}

\end{document}